\pdfoutput=1

\documentclass[11pt]{article}

\usepackage{PRIMEarxiv}

\usepackage{float}
\usepackage{times}
\usepackage{latexsym}
\usepackage{url}
\usepackage{booktabs}
\usepackage{longtable}
\usepackage{listings}
\usepackage{xcolor}
\usepackage{setspace}
\usepackage{etoolbox}
\usepackage{parskip}

\AtBeginEnvironment{table}{\vspace{6pt}}
\AtEndEnvironment{table}{\vspace{6pt}}

\usepackage[T1]{fontenc}

\usepackage[utf8]{inputenc}

\usepackage{microtype}

\usepackage{inconsolata}

\usepackage{graphicx}

\title{Querying Databases with Function Calling}

\author{
    Connor Shorten \\
    Weaviate \\
    \And
    Charles Pierse \\
    Weaviate \\
    \And
    Thomas Benjamin Smith \\
    Weaviate \\
    \And
    Karel D'Oosterlinck \\
    Contextual AI \\
    \And
    Tuana Celik \\
    Weaviate \\
    \And
    Erika Cardenas \\
    Weaviate \\
    \And
    Leonie Monigatti \\
    Weaviate \\
    \And
    Mohd Shukri Hasan \\
    Weaviate \\
    \And
    Edward Schmuhl \\
    Weaviate \\
    \And
    Daniel Williams \\
    Weaviate \\
    \And
    Aravind Kesiraju \\
    Morningstar \\
    \And
    Bob van Luijt \\
    Weaviate \\
}

\begin{document}
\maketitle

\begin{abstract}
The capabilities of Large Language Models (LLMs) are rapidly accelerating largely thanks to their integration with external tools. Querying databases is among the most effective of these integrations, enabling LLMs to access private or continually updating data. While Function Calling is the most common method for interfacing external tools to LLMs, its application to database querying as a tool has been underexplored. In this report, we propose and extensively test a tool definition for database querying that unifies accessing data with search queries, filters, or a combination both, as well as transforming results with aggregation and groupby operators. Our proposed tool definition additionally enables the LLM to route queries across multiple collections of data. To evaluate its effectiveness, we conduct a study with 8 LLMs spanning 5 model families. We present a novel pipeline adapting the Gorilla LLM framework to create synthetic search database schemas and queries. We present an analysis comparing our proposed DBGorilla dataset to popular text-to-SQL benchmarks such as BIRD, Spider, and WikiSQL. Using the DBGorilla benchmark, we show that Claude 3.5 Sonnet, GPT-4o, GPT-4o mini, and Gemini 1.5 Pro are all highly effective at utilizing our proposed tool definition for querying databases. We primarily evaluate these models with the Exact Match of predicted and ground truth query APIs. To gain a more holistic understanding of model performance, we also report Abstract Syntax Tree (AST) alignment scores and LLM-as-Judge preference rankings of predicted queries. Among the eight models tested, Claude 3.5 Sonnet achieves the highest performance with an Exact Match score of 74.3\%, followed by GPT-4o mini at 73.7\%, GPT-4o at 71.8\%, and Gemini 1.5 Pro at 70.2\%. We further breakdown these results by API component, finding that LLMs are highly effective at utilizing operators on boolean-valued properties, but struggle to understand text property filters and differentiate them from search queries. We further visualize the performance across the synthetic use cases, showing robust results with the higher performing models such as GPT-4o, but significant performance variance across use cases from lower performing models. To further understand the impact of tool definitions on connecting LLMs with querying databases, we conduct ablation studies exploring the impact of parallel tool calling, adding a rationale as an argument of the tool call, using a separate tool per database collection, and tool calling with structured outputs. We find minimal performance variance across these ablation experiments with GPT-4o. Our findings demonstrate the effectiveness of enabling LLMs to query databases with Function Calling. We have open-sourced our experimental code and results at github.com/weaviate/gorilla.
\end{abstract}

\section{Introduction}

Large Language Models (LLMs) have achieved remarkable successes in natural language understanding and reasoning. The applications of LLMs are rapidly advancing as they are connected with other software tools in architectures broadly described as Compound AI Systems. From Zaharia et al., a Compound AI System "tackles AI tasks using multiple interacting components, including multiple calls to models, retrievers, or external tools" \cite{CompoundAISystems}. Connecting AI models to external software tools complements their weaknesses, such as accessing private or continually updating data, as well as symbolic computation. However, understanding the most effective interface between LLMs and external tools remains an open question. Pioneered by works such as ReAct \cite{react} and the Gorilla LLM \cite{gorilla}, Function Calling has emerged as a powerful architecture for Compound AI Systems. Defined in OpenAI's developer documentation, "function calling enables developers to connect language models to external data and systems. You can define a set of functions as tools that the model has access to, and it can use them when appropriate based on the conversation history. You can then execute those functions on the application side, and provide results back to the model"\cite{OpenAIFunctionCalling}. OpenAI, as well as many other model providers and open-source tools provide JSON schemas with which to define these functions with. In our work, we propose and test a function definition following this interface for querying databases.

\begin{figure*}
    \centering
    \includegraphics[width=1\linewidth]{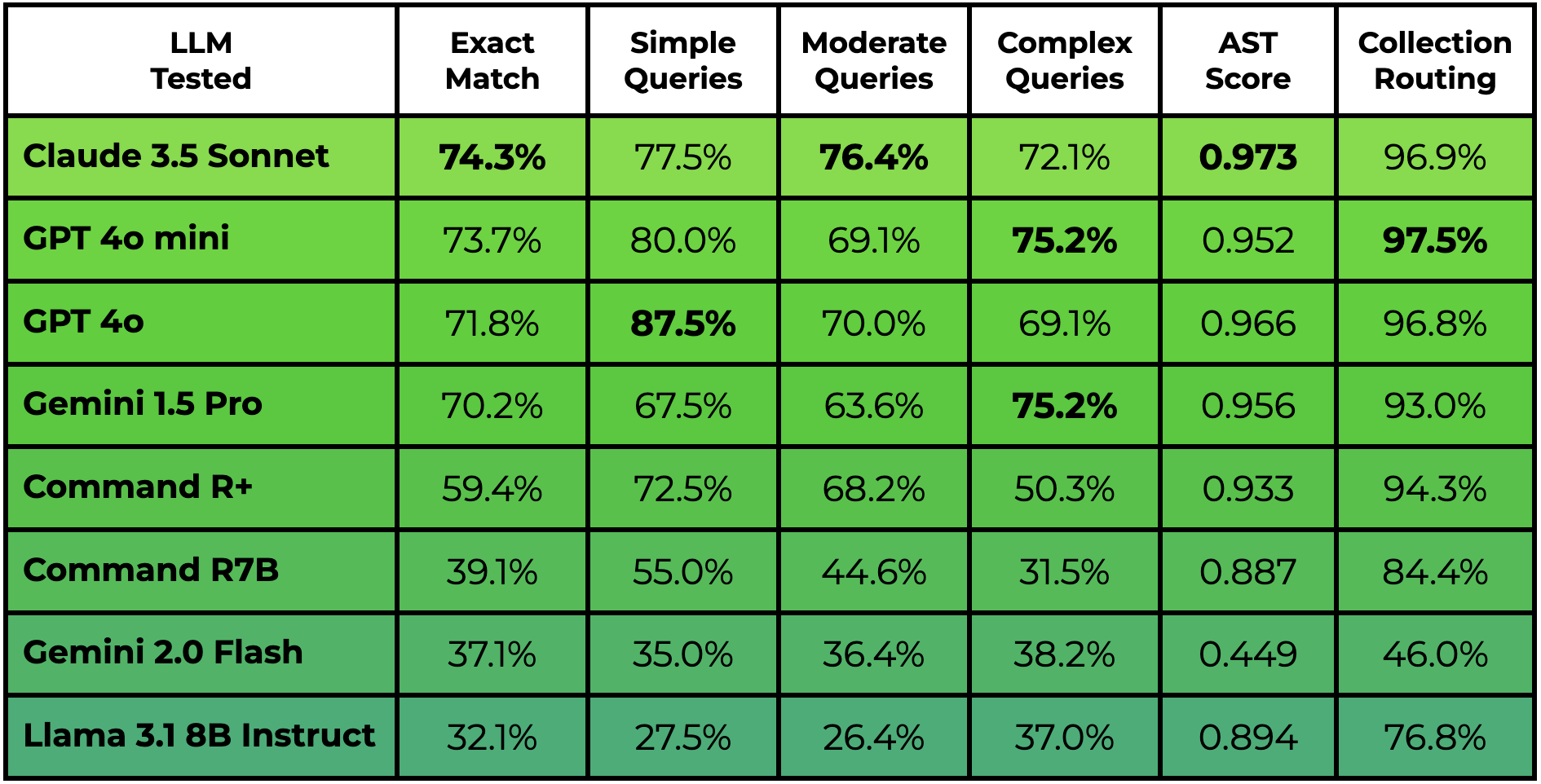}
    \caption{DBGorilla Leaderboard results (last updated January 1st, 2025). The Exact Match and AST Score columns report the respective averages across all tested queries. Query scores are further separated into categories of "Simple", "Moderate", and "Complex" according to how many arguments are used in the ground truth function call with 1, 2, and 3 or more, respectively. Collection routing reports the percentage the predicted query is routed to the correct database collection. }
    \label{fig:enter-label}
\end{figure*}

Function Calling has seen enormous application in Compound AI System design, but has mostly been limited to relatively simple tools. For example, a \textbf{get\_unread\_emails} function that does not require any input arguments and returns unread emails. Another common example demonstrating Function Calling is the \textbf{get\_weather} function. Interestingly, this function has constraints on its input argument, requiring a 5 digit integer-valued zip code to access the weather across cities in the United States of America. This is very similar to how one might approach interfacing database querying with SQL through Function Calling, requiring constraints on a string-valued \textbf{sql\_query} argument of a \textbf{query\_database} function. Unfortunately, Function Calling does not yet support advanced constraints on input arguments, limiting the effectiveness of SQL with Function Calling. We instead show how we can decompose query APIs into a series of optional JSON-valued arguments to better utilize SQL-style query operators with Function Calling.

Most previous works on database querying with machine learning models has focused on text-to-SQL translations. However, as highlighted in recent works such as Spider 2.0, "real-world data are stored across a diverse array of database systems, each with its own unique SQL dialects, introducing a wide range of SQL syntax and functions" \cite{spider2}. These SQL dialects are often highly specific to the underlying database system being queried. For example, the query operators available in a database system built on top of the relational data model differ from document or graph data models. Thus we propose disentangling the SQL syntax from the particular query operators the LLM has access to. Shown in Figure 2, we translate natural language commands into Function Calling arguments, instead of SQL. One notable benefit of decoupling query operators from SQL syntax is how easily we can plug and play with different query operators. As one example of what this allows us to do, we unify structured data access and result transformations with search queries. As highlighted by Wu et al. in the presentation of the STaRK benchmark, "many previous works studied textual and relational retrieval tasks as separate topics" \cite{STaRK}. We propose that this isolation is due to textual and relational retrieval tasks being built on distinct underlying data models. The relational data model typically assumes multiple normalized tables linked together with foreign keys and frequent use of the JOIN query operator. On the other hand, textual retrieval typically assumes a single collection of data with a single text property per object that is stored in a search index. Our tool definition for Function Calling adds the search operator to a collection of operators derived from the relational data model. This tool definition can be trivially mapped to and from custom SQL dialects or extended with functionality from less conventional data models such as SPARQL \cite{SPARQL}. We can also easily integrate new operators introduced in emerging languages such as LOTUS \cite{LOTUS}, TAG \cite{TAG}, or SUQL \cite{SUQL} to Function Calling arguments.

To evaluate LLMs' ability to format database queries with Function Calling, we present the DBGorilla benchmark. DBGorilla is an adaption of the Berkley Function Calling Leaderboard \cite{BFCL} and Gorilla LLM following the use of Self-Instruct to create synthetic natural language commands targeted towards the combinatorics of API operators. Extending the original Gorilla LLM methodology, DBGorilla requires a synthetic database schema in order to ground the natural language commands and ground truth query operator values. We thus begin by presenting a framework to generate synthetic database use cases. Each generated use case represents a distinct business domain and consists of three interrelated collection schemas. Every collection is designed with a searchable text property and three additional properties, one numeric, one textual, and one boolean, to enable comprehensive testing of different query patterns. This structured approach allows us to systematically assess how well LLMs can interpret database schemas and translate natural language requests into appropriate database operations. Given this dataset of schemas, we then create a comprehensive test dataset of queries covering all combinations of query operators defined in the tool schema. These capabilities include search queries for finding relevant results based on relevance ranking algorithms, property filters for matching on integer, text, and boolean fields, aggregations for computing statistics over integer, text, and boolean properties, and grouping operations to segment results by property values.

Utilizing the DBGorilla benchmark, we compare 8 LLMs from 5 model families on the task of choosing the correct database query API given the schemas of collections available to query and a natural language command as input. Some queries only require a single API, such as: \textbf{How many unique menu items are priced under \$20?}. This can be answered with an integer filter that sets the price less than \$20. Contrastively, the query: \textbf{What is the average price of seasonal specialty menu items under \$20, grouped by whether they are vegetarian or not?} requires a search query for "seasonal specialties", setting a price filter of less than \$20, calculating the average price of the results, and grouping them by the isVegetarian boolean property. All queries require routing to the appropriate collection.

\begin{figure*}
    \centering
    \includegraphics[width=0.8\linewidth]{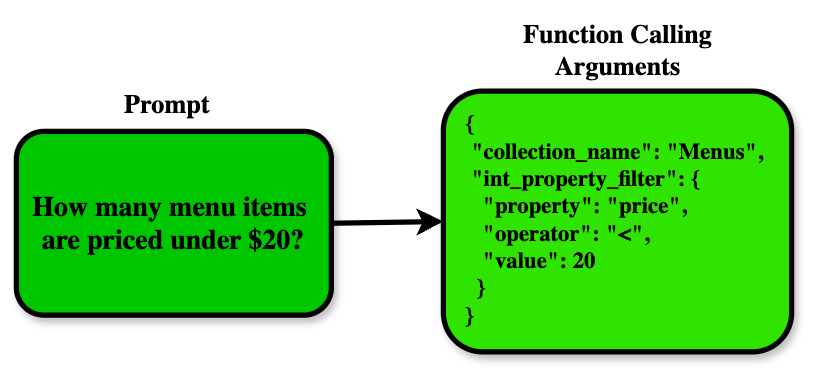}
    \caption{An illustration of a natural language command, \textbf{How many menu items are priced under 20?}, translated to Function Calling arguments for database querying.}
    \label{fig:enter-label}
\end{figure*}

We primarly report Exact Match as our performance metric. Exact Match is a boolean metric assessing if the predicted query from the LLM is identical to the ground truth query. This metric is particularly insightful thanks to the targeted nature of the synthetic benchmark dataset. To gain more insight into the accuracy of predicted queries, we also report the structural similarity of predicted and ground truth queries with Abstract Syntax Tree (AST) scoring. The AST approach breaks down queries into their hierarchical components starting with the target collection as the root node, followed by branches for search queries, filters, aggregations, and grouping operations. The scoring heavily weights getting the target collection correct (40\% of total score), as this is fundamental to query correctness. The remaining score is evenly distributed (15\% each) across matching the search query text, filter specifications, aggregation operations, and grouping property. This hierarchical scoring approach allows us to quantify partial successes in query generation and identify specific areas where models struggle. We additionally present an LLM-as-judge preference ranking analysis as another lens into LLM querying performance. This entails using an LLM to rank the predicted queries from each of the 8 LLMs for a given natural language command and database schema. We further present the Collection Routing accuracy, the percentage of time the predicted query targets the correct collection. Finally, we report the tool selection rate as the number of times the LLM decides it needs to perform a function call in the initial step of the Function Calling framework.

In summary our contributions are as follows:
\begin{itemize}
    \item We introduce DBGorilla, a collection of 5 use cases, each with 3 related collections and 4 properties per collection. We additionally present 315 queries, 63 unique combinations of query APIs for each of the 5 use cases. We present a cost analysis of maintaining and creating this benchmark, as well as a discussion of directions for expansion.
    \item We present a tool definition schema that unifies search queries and structured data access. We demonstrate that Claude 3.5 Sonnet, GPT-4o, GPT-4o-mini, and Gemini 1.5 Pro are all highly effective at formatting API calls for querying a search database using this tool definition with Function Calling.
    \item We present ablation studies demonstrating that parallel tool calling, rationale generation, separate tool definitions per collection, and tool calling with structured outputs all have minimal impact on the resulting performance.
\end{itemize}

\section{Related Works}

\subsection{Compound AI Systems}

Tool use is one of the most promising opportunities to improve the capabilities of LLMs. There are two common design patterns for interfacing tool use in Compound AI Systems: Function Calling and Flow Engineering \cite{alphacodium}. Visualized in Figure 4, Function Calling entails equipping the LLM with a set of functions described in the prompt. The LLM inference is then orchestrated in a function calling loop. At each step, the LLM either chooses to complete the response, or call one or multiple functions and wait for their respective responses to continue the next iteration of the loop. Contrastively, Flow Engineering describes a pre-determined flow of inferences and external tools calls. This abstraction helps clarify how tools are interfaced to LLMs. However, there is a significant overlap and this is a constantly evolving area of AI research. For example, an engineered LLM and tool calling flow could be itself abstracted and interfaced as a function for the agent to call. In a similar analog, a flow could implement the open-ended looping core to the definition of Function Calling. Understanding these distinctions is important for the evolution of prior works on interfacing search and database querying as an LLM tool. In either case, we need methods to evaluate how well LLMs can select the correct tool for the task and format the tool's respective arguments \cite{gorilla}.

Search has been one of the most commonly used tools for LLMs. Most commonly, this has taken the shape of RAG \cite{RAG}, a pre-determined flow of retrieval with the user input as query, followed by response generation. RAG flows were further pioneered with architectures such as Baleen RAG, in which the user input is first translated into search queries with an LLM inference, sent to a retrieval engine, and passed into a final response generation. One of the early efforts to expand search to the Function Calling interface was WebGPT \cite{WebGPT}, in which the LLM can format search queries to send to the web, as well as paginate through the results. Zhang et al. debuted the term “Agentic Information Retrieval” \cite{agenticir} to capture the intersection of learning to search with the Function Calling interface.

\subsection{Text-to-SQL}

Developing mostly in parallel to search as a tool, AI researchers and practitioners have been exploring the use of database APIs as a tool. Even before breakthrough capabilities in LLMs, Text-to-SQL research has been a heavily studied discipline \cite{texttosqlsurvey}. Text-to-SQL research has mostly targeted the application of making it easier for humans to learn how to query databases. We primarily studied three popular Text-to-SQL benchmarks in this work: WikiSQL \cite{wikisql}, Spider \cite{spider1, spider2}, and BIRD \cite{birdsql}. WikiSQL consists of 80,654 hand-annotated examples of questions and SQL queries distributed across 24,241 tables from Wikipedia. The original Spider dataset contains 10,181 questions and 5,693 unique complex SQL queries on 200 databases with multiple tables, covering 138 different domains. BIRD contains 12,751 Text-to-SQL pairs and 95 databases spanning 37 professional domains. We visualize samples from the BIRD dataset in Figure 3 to help readers further understand the current state-of-the-art in Text-to-SQL benchmarking. In Spider 2.0, the authors diverge from Text-to-SQL prediction to Text-to-SQL workflows, adopting a more holistic view of data querying and transformation with SQL.

Now that most databases are evolving to support search indexes and integration with LLMs, additional query languages are emerging to expand SQL, such as LOTUS \cite{LOTUS} and SUQL \cite{SUQL}. Our work is further related to managing multiple database collections in architectures such as Data Warehouses, Lakehouses \cite{lakehouse}, or Ontologies \cite{ontology}. In order to study machine learning for databases, we need new benchmarks and datasets reflective of the challenges of database systems. Similarly to our synthetic schemas, Lim et al. \cite{dbgyms} present Database Gyms, focusing on system-level optimization and workload simulation.

\begin{figure*}
  \centering
  \includegraphics[width=0.8\linewidth]{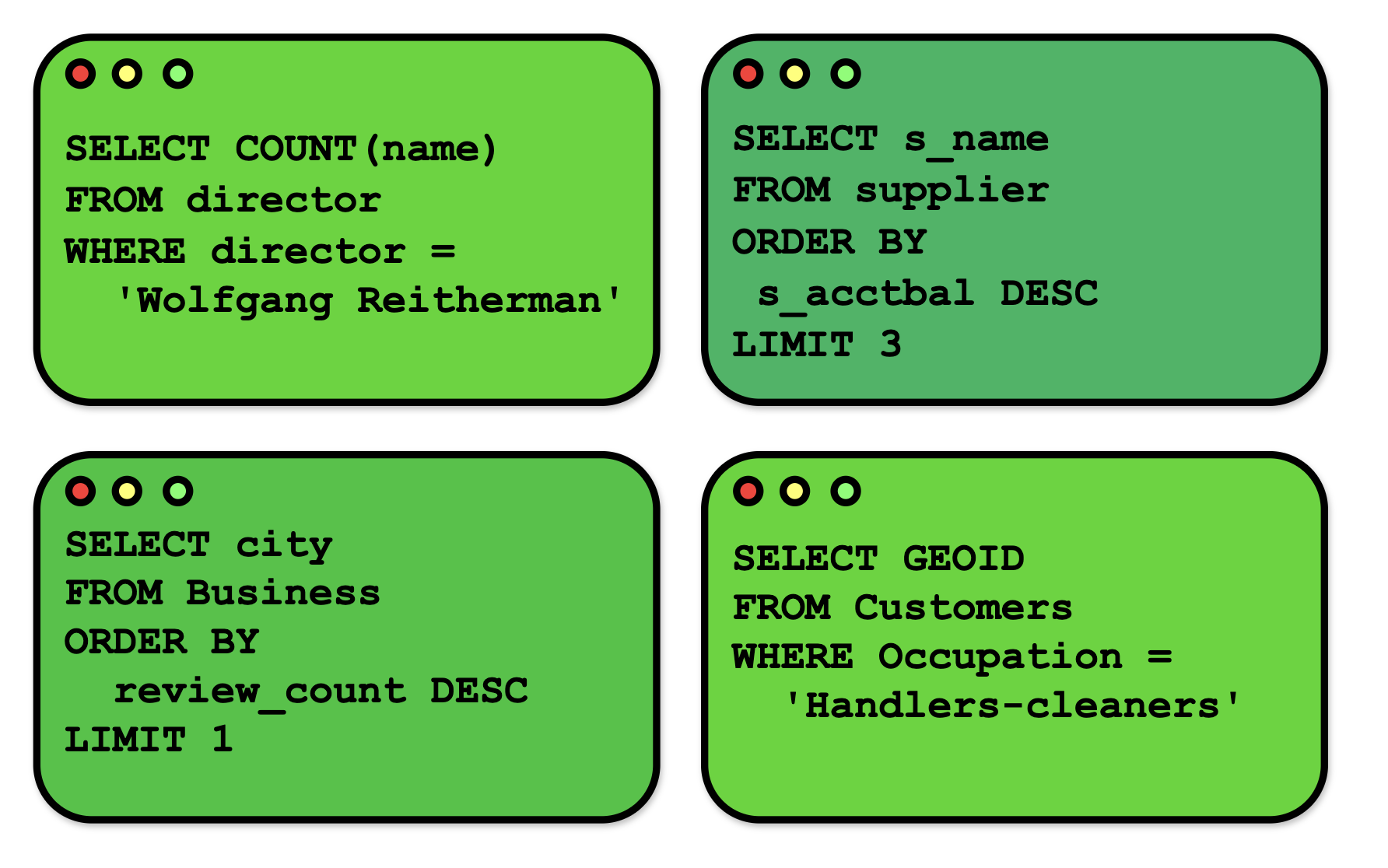}
    \caption{Examples of queries in the BIRD Text-to-SQL benchmark \cite{birdsql}. We visualize these to help readers gain a better understanding of how Text-to-SQL is currently studied and how BIRD differs from DBGorilla.}
  \label{fig:enter-label}
\end{figure*}

\section{Methodology}

\subsection{Details of Function Calling Setup}

As described in the context of Compound AI Systems, Function Calling entails equipping the LLM with a set of functions described in the prompt. The LLM inference is then orchestrated in a Function Calling loop. At each step, the LLM either chooses to complete the response, or call one or multiple functions and wait for their respective responses to continue the next iteration of the loop. We limit the Function Calling loop to a single step and evaluate the accuracy of the predicted Function Calling arguments. We use the tool definition shown in Appendix A.

\begin{figure*}
    \centering
    \includegraphics[width=0.8\linewidth]{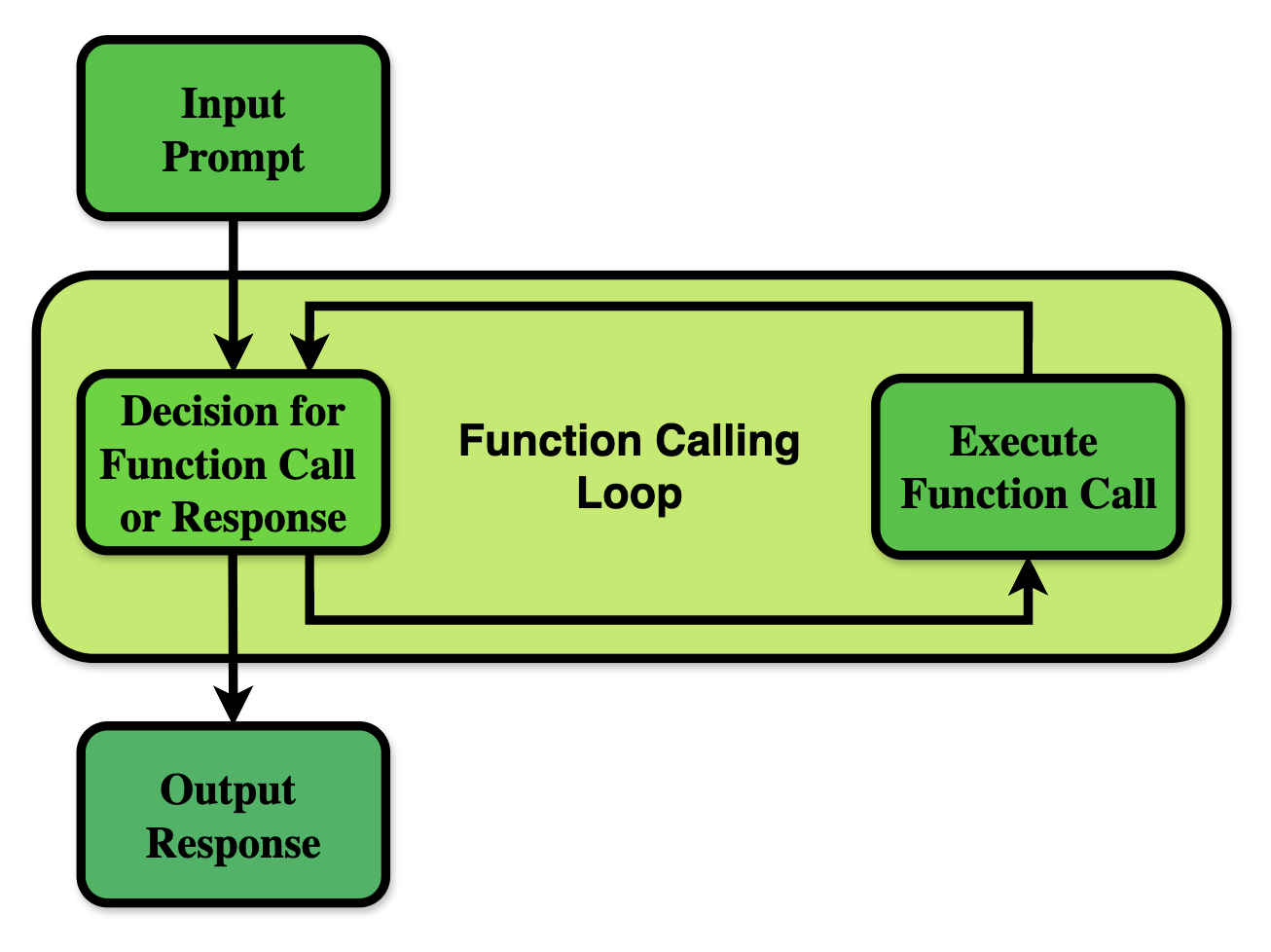}
    \caption{An illustration of the Function Calling loop. Beginning with the user's input prompt, the LLM then enters a loop where it can either choose to call one or multiple functions, or return a response to the user. If a function is called, the function is executed, the response is sent back to the LLM, and the Function Calling loop continues.}
    \label{fig:enter-label}
\end{figure*}

More concretely, we store each record in our DBGorilla dataset with the properties, \textbf{nl\_command}, \textbf{ground\_truth\_query}, and \textbf{schema}. We send the schema to the database to create the collections. We then retrieve metadata about available collections and their properties from the database, although you could also achieve this from the schema stored in the dataset. We then parse this meta information into a description string detailing the collections and their properties, as well as a list of collection names that are used for routing queries as an enum-valued API argument. The description of collections and database querying is carefully constructed to fit within 1024 tokens due to token limits from the LLMs tested when using their respective Function Calling SDKs. The tool schema then exposes a \textbf{query\_database} function with parameters tailored to the database's querying capabilities. The \textbf{collection\_name} argument is the only required argument, which is restricted to the enumerated list of available collections. Optional parameters enable search queries, filters, aggregations, and groupby. We then pass the natural language command stored in the dataset to the LLM with the tool definition and record the arguments used in the function call. If the LLM chooses not to call a function, it achieves a score of 0 for this instance. This is motivated by the highly targeted nature of these natural language commands, which we will discuss further later on. We test the GPT-4o and GPT-4o mini LLMs \cite{GPT4}, Gemini 1.5 Pro and Gemini 2.0 Flash experimental \cite{Gemini}, Claude 3.5 Sonnet \cite{ClaudeSonnet}, Command R+ and Command R7B \cite{CommandRDocs}, and Llama 3.1 8B Instruct \cite{Llama}.

\subsection{DBGorilla Dataset Construction}

We present a novel dataset for measuring the effectiveness of LLMs to query databases with Function Calling, DBGorilla. DBGorilla consists of two phases: synthetic schema and query generation. The construction of this dataset heavily relies on structured generation methods \cite{outlines, LetMeSpeak, StructuredRAG}. Structured outputs lets us easily control the validitiy of generated schema and query structure. We can easily use this framework to generate more synthetic schemas, or schemas with different property distributions. Further, we can easily switch out the query operators available to the LLM and construct a corresponding query set. In our discussion section, we present further thoughts on how to extend this benchmark and predictions for the evolution of benchmarking text-to-SQL and database use in Compound AI Systems.

\subsubsection{Synthetic Database Schemas}

The schema generation process utilizes GPT-4o to create synthetic database schemas through a structured prompt. The database generation prompt contains a reference example schema that demonstrates the desired structure and detail level, along with specific requirements for each collection including two text properties (one with rich searchable content), one numeric property, and one boolean property. We note that these requirements can be varied to create diverse schema types, such as eight boolean properties and  one searchable text property if desired. In order to test routing queries to multiple collections, the generator is further instructed to ensure collections are meaningfully related. However, we do not explicitly link these collections together with foreign key relationships. Using these inputs, we produce five schema sets, each containing three interconnected collections. An example is shown in Table 1, a use case modeling a restaurant system with Restaurants, Menus, and Reservations collections. Each generated schema follows consistent conventions with collection names in camel case format, comprehensive property definitions including types and detailed descriptions. We further generate a \textbf{use\_case\_overview} to facilitate interpretability of generated schemas.

\begin{table}[htbp]
\centering
\setlength{\tabcolsep}{6pt} 
\renewcommand{\arraystretch}{1.1} 
\resizebox{0.5\textwidth}{!}{ 
\begin{tabular}{ll}
\toprule
\textbf{Collection Name} & \textbf{Property Name} \\
\midrule
Restaurants & name (string) \\
            & description (string) \\
            & averageRating (number) \\
            & openNow (boolean) \\
\midrule
Menus       & menuItem (string) \\
            & itemDescription (string) \\
            & price (number) \\
            & isVegetarian (boolean) \\
\midrule
Reservations & reservationName (string) \\
             & notes (string) \\
             & partySize (number) \\
             & confirmed (boolean) \\
\bottomrule
\end{tabular}
}
\vspace{1em}
\caption{A visualization of the Restaurant synthetic database schema.}
\label{tab:weaviate-collections}
\end{table}

\subsubsection{Synthetic Queries}

The query generation process follows the algorithms introduced in Self-Instruct \cite{selfinstruct} to create comprehensive test cases of API use cases. We extend Self-Instruct to add Reflexion \cite{15} to assess, and potentially correct, generated queries with another LLM inference. The addition of Reflexion to synthetic query generation helps us get a quantitative sense of dataset quality and qualitatively when manually inspecting individual queries with the user interface shown in Figure 6. We generate all valid combinations of query operators, including options for semantic search, filters (integer, text, boolean), aggregations, and grouping operations. For each operator combination, we create a Pydantic model that specifies required fields and includes descriptions of how each operator should be used, ensuring that the natural language query necessitates all selected operators. Using GPT-4o, we then generate natural language queries by providing the database schema and operator requirements, validating that each query requires all specified operators to be answered correctly by creating structured output models on the fly for each query combination. We run this process to yield 63 queries per schema. We produce 315 queries in total across 5 database schemas.  We use our dataset visualizer GUI shown in Figure 6 to manually verify the quality of these queries.

\subsection{Evaluating Predicted Queries}

We present three strategies for evaluating the quality of predicted database queries with Function Calling. We primarily use Exact Match Scoring evaluation. Exact Match scoring returns a boolean assessment if the predicted and ground truth queries are exactly identical. We additionally utilize Abstract Syntax Tree (AST) evaluation. AST scores are a highly effective method to measure how aligned predicted queries are with the ground truth API components the natural language command is crafted to target. We further introduce a preference ranking evaluation using LLM-as-judge. This is largely inspired by the challenge of evaluating real-world queries that do not come with a ground truth API path. Preference ranking evaluation further offers additionally flexibility in query quality judgement and leniency in cases where the LLM chooses not to call a function. Although we note that due to the highly targeted nature of natural language commands in our dataset, it is never effective to choose not to call a function. Additionally, in our controlled environment with ground truth API queries, we can gain insight into the alignment of these metrics.

The AST evaluation methodology employs a weighted scoring system to assess query similarity. The largest weight (40\%) is assigned to correctly matching the target collection, mismatching collections results in a score of 0. The remaining 60\% is evenly distributed (15\% each) across four components: search queries, filters, aggregations, and group by operations. We do not evaluate if the search queries are similar, only if the predicted and ground truth queries both use the search query or not. Contrastively, filters, aggregations, and groupby values must be identical to the ground truth query to achieve the 0.15 points for the match. The final score ranges from 0.0 to 1.0, with 1.0 indicating perfect structural alignment across all elements.

In order to better understand the performance of Large Language Models to format database queries, we conduct an llm-as-judge preference ranking test. The test utilized structured outputs to rank the 8 LLM responses on a scale of 1 to 8. The model additionally presents an explanation of why it decided on the particular ranking for the user query. We report the number of 1st place rankings each model receive, as well as a weighted score across ranks. We report the number of first place ranks each model achieves, as well as a weighted rank scoring. The weighted scoring system grants 100 points for first place, 70 for second, 50 for third, 35 for fourth, 25 for fifth, 20 for sixth, 15 for seventh, 10 for eighth, 5 for ninth, and 0 for tenth place and beyond. This approach heavily rewards finishing near the top but still provides partial credit for mid-range positions, aiming to capture strong overall performance rather than sporadic high placements.

\subsection{DBGorilla Expansion and Maintenance Cost}

The computational costs associated with running the DBGorilla benchmark reveal significant economic implications for model selection, deployment, and benchmark maintenance. As shown in Table 2, there is a stark contrast in costs across model families, with Claude 3.5 Sonnet being the most expensive at \$2.84 total cost, while Command R7B is the most economical at \$0.03. These cost differentials do not strictly correlate with performance. For example, GPT-4o mini achieves strong results (73.7\% Exact Match score) at \$0.12 total cost, suggesting a pareto-optimal frontier of price-performance. The benchmark's total token consumption (245,000 input tokens and 140,000 output tokens) provides a standardized basis for comparing model costs.
From a benchmark maintenance perspective, the aggregate cost to evaluate all eight models is approximately \$8.10 per run. This translates to roughly \$97.20 annually for monthly evaluations. These maintenance costs are particularly relevant for the research community, as they enable regular updates to track the rapid evolution of LLM capabilities in database querying. The total cost supports the DBGorilla benchmark's sustainability and encourages broader participation in model evaluation and provide transparency about the resources required for replication studies. Generating the synthetic benchmark of 315 queries required a total of 413,516 input tokens and 86,457 output tokens. Shown in Table 2, this totals to \$1.89 with OpenAI GPT-4o.

\begin{table}[h!]
\centering
\resizebox{1.0\linewidth}{!}{
\begin{tabular}{lccccc}
\toprule
\textbf{Model} & \textbf{Input Cost (\$)} & \textbf{Output Cost (\$)} & \textbf{Total Cost (\$)} & \textbf{Input Pricing (\$/1M)} & \textbf{Output Pricing (\$/1M)} \\
\midrule
Claude 3.5 Sonnet      & 0.74   & 2.10   & 2.84   & 3.00   & 15.00   \\
OpenAI GPT-4o          & 0.61   & 1.40   & 2.00   & 2.50   & 10.00   \\
Command R+             & 0.61   & 1.40   & 2.00   & 2.50   & 10.00   \\
Gemini 1.5 Pro         & 0.31   & 0.70   & 1.01   & 1.25   & 5.00    \\
GPT-4o Mini            & 0.04   & 0.08   & 0.12   & 0.15   & 0.60    \\
Llama 3.1 8B Instruct  & 0.02   & 0.014  & 0.04   & 0.10   & 0.10    \\
Gemini 1.5 Flash       & 0.02   & 0.04   & 0.06   & 0.075  & 0.30    \\
Command R7B            & 0.01   & 0.02   & 0.03   & 0.0375 & 0.15    \\
\bottomrule
\end{tabular}%
}
\vspace{1em}
\caption{Cost comparison of models tested (last updated January 1st, 2025).}
\label{tab:cost-comparison}
\end{table}

\section{Experimental Results}

The DBGorilla leaderboard shown in Figure 1 illustrates a clear hierarchy in model performance across different evaluation metrics, with interesting patterns in performance. Claude 3.5 Sonnet leads overall with an Exact Match score of 74.3\%, followed closely by GPT-4o mini at 73.7\%, GPT-4o at 71.8\% and Gemini 1.5 Pro at 70.2\%. There is then a somewhat steep dropoff to 59.4\% from Command R+, followed by a significant performance gap to Gemini 2.0 Flash (exp) at 37.1\% and Llama 3.1 8B Instruct at 32.1\%. Performance varies across different query complexities. For simple queries (requiring a single argument), the top models perform remarkably well, with GPT-4o achieving a score of 87.5\% and Claude Sonnet 3.5 reaching 77.5\%. Looking across query complexities, measured as requiring more than 1 operator, we see an encouraging robustness in performance. Claude 3.5 Sonnet's performance on Simple Queries at 77.5\% is not too far off its' effectiveness with Complex Queries at 72.1\%. The collection routing metric reveals another interesting pattern, while most top models hover around 96 to 98\%, Command R+ stands out with 94.3\% accuracy despite a relatively lower 59.4\% total Exact Match score, suggesting it has a particular strength in understanding and correctly selecting the appropriate database collection. The Abstract Syntax Tree (AST) scoring analysis reveals further nuance into model performance beyond Exact Match metrics. Claude 3.5 Sonnet achieves a nearly perfect 0.973 AST score. This indicates near perfect structural understanding of query components in cases missing the strict criterion of Exact Match scoring. The top four performing models all maintain AST scores above 0.95, additionally illustrating strong comprehension of query operator structure.
 
\subsection{Component and Schema Variance Analysis}

To better understand where the LLMs went wrong in their predicted queries, we break performance down by API component involved in the ground truth queries. We present a radar plot visualization of this in Figure 5 and a detailed view of results in Table 6 and Table 7. Boolean filters stand out as the most successfully handled component across all models, with GPT-4o and Claud 3.5 Sonnet both achieving 87.5\% Exact Match accuracy. However, their performance drops on boolean aggregations with scores of 62.5\% and 66.25\%, respectively. Most interestingly, the models show a significant performance decline on text filters, failing to distinguish them from search queries. The evaluation across different database schemas, shown in Table 2, reveals varying levels of domain adaptability among the models. GPT-4o demonstrated the most consistent cross-domain performance, with results ranging from 73.44\% on the Restaurants use case to 67.8\% on the Visual Arts use case, a range of 5.64\%. This stability stands in marked contrast to smaller models. Gemini 2.0 Flash exhibited dramatic performance variance, ranging from 57.81\% to 23.44\%. These findings indicate a strong correlation between model size and the ability to maintain consistent performance across varied domains, with larger models demonstrating superior schema adaptability.

\begin{figure*}
    \centering
    \includegraphics[width=1\linewidth]{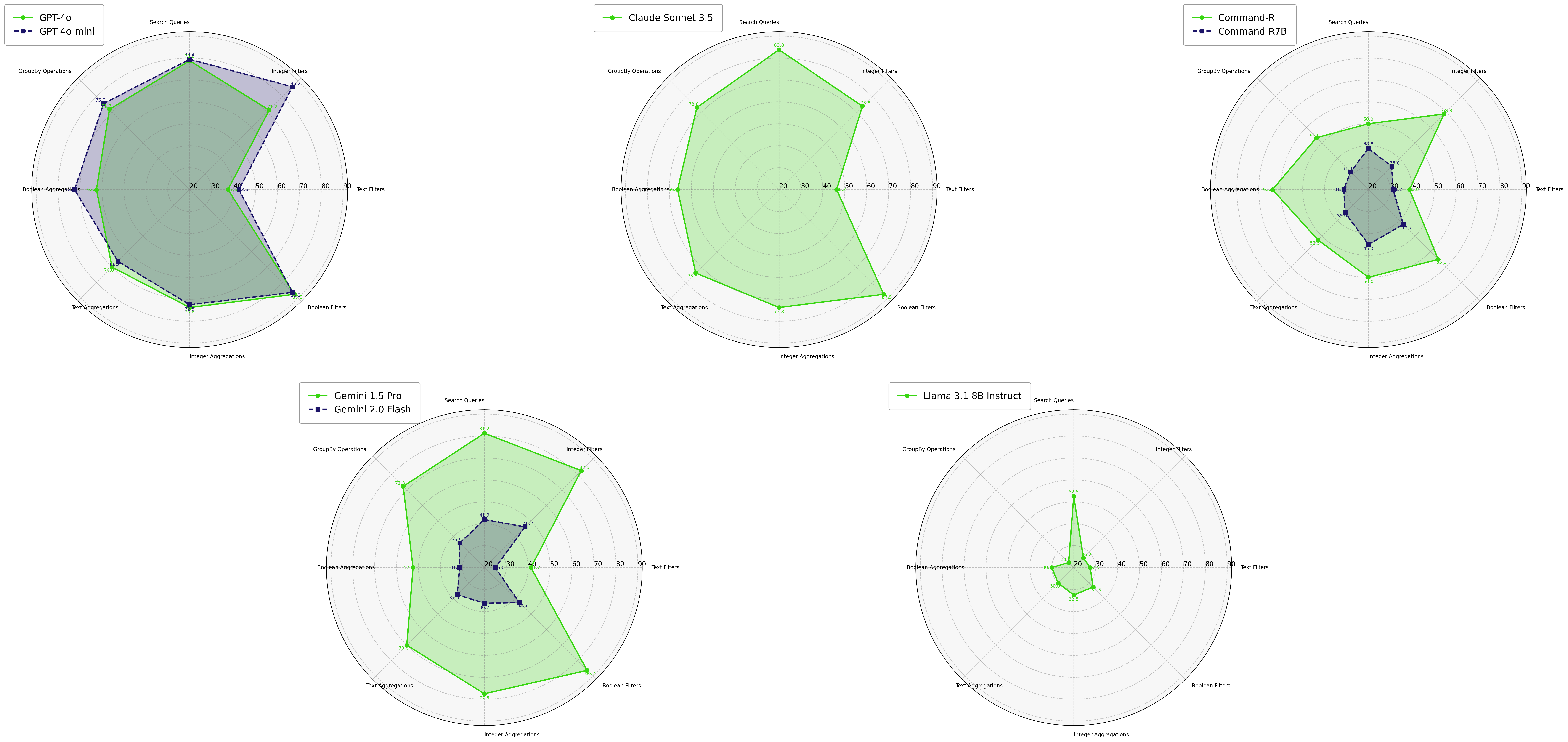}
    \caption{Radar plots highlighting how well each model tested can access particular Search Database API components.}
    \label{fig:enter-label}
\end{figure*}

\begin{table}[h]
\centering
\setlength{\tabcolsep}{10pt}
\renewcommand{\arraystretch}{1.2}
\resizebox{\textwidth}{!}{%
\begin{tabular}{lccccc}
\toprule
\textbf{Model} & \textbf{Restaurants} & \textbf{Health Clinics} & \textbf{Courses} & \textbf{Travel Planning} & \textbf{Visual Art} \\
\midrule
GPT-4o            & 73.44\% & 76.56\% & 70.31\% & 70.31\% & 67.80\% \\
GPT-4o mini       & 75.00\% & 75.00\% & 68.75\% & 75.00\% & 74.58\% \\
Claude 3.5 Sonnet & 71.88\% & 73.44\% & 71.88\% & 71.88\% & 83.05\% \\
Command R+        & 60.94\% & 50.00\% & 54.69\% & 60.94\% & 71.19\% \\
Command R 7B      & 39.06\% & 37.50\% & 35.94\% & 43.75\% & 38.98\% \\
Gemini 1.5 Pro    & 73.44\% & 65.62\% & 68.75\% & 73.44\% & 69.49\% \\
Gemini 2.0 Flash  & 57.81\% & 23.44\% & 35.94\% & 25.00\% & 44.07\% \\
Llama 3.1 8B Instruct      & 31.25\% & 37.50\% & 31.25\% & 28.12\% & 32.20\% \\
\bottomrule
\end{tabular}
}
\vspace{1em}
\caption{Performance Across Different Schemas for All Tested Models}
\label{tab:schema-performance}
\end{table}

\subsection{No Tool Selected}

Shown in Figure 4, the Function Calling Loop begins with an initial design to call a function or respond to the user. We find that all LLMs tested occasionally skip function calling and immediately return the response without querying the database. This is particularly emphasized in the Gemini 2.0 Flash model, which skips function calling more often than not at a rate of 53.97\%. This also explains Gemini 2.0 Flash's poor performance on the broader set of evaluation metrics such as Exact Match, Abstract Syntax Scoring, and Preference Ranking.

\begin{table}[h]
\centering
\setlength{\tabcolsep}{7pt}
\renewcommand{\arraystretch}{1.1}
\begin{tabular}{lr}
\toprule
\textbf{Model} & \textbf{No Tool Selected Rate (\%)} \\
\midrule
GPT-4o & 2.86\% \\
GPT-4o-mini & 0.95\% \\
Claude 3.5 Sonnet & 3.17\% \\
Command R+ & 4.44\% \\
Command R7B & 13.33\% \\
Gemini 1.5 Pro & 5.40\% \\
Gemini 2.0 Flash & \textbf{53.97\%} \\
Llama 3.1 8B & \textbf{21.90\%} \\
\bottomrule
\end{tabular}
\vspace{1em}
\caption{As shown in Figure 4, the LLM is tasked with the decision to call a function or submit a response to the user. Despite our benchmark being particularly constructed to create natural language commands that require database queries, some models choose not to call a function in a surprisingly high percentage of test cases.}
\label{tab:failure-rate}
\end{table}

\subsection{Preference Rankings}
Across the 315 tested queries, only 5 result in identical predictions for the 8 LLMs tested. On average, each query has 5.8 unique predictions from the 8 LLMs. From the preference ranking results shown in Table 3, Gemini 1.5 Pro, GPT-4o mini and GPT-4o emerge as the most favored models, consistently occupying the highest portion of first-place votes. Meanwhile, Command R7B, Llama 3.1 8B Instruct, and Claude 3.5 Sonnet often fall toward the bottom in aggregated rankings. These results vary significantly from the Abstract Syntax Tree (AST) evaluations, which highlighted the same top three as highly skilled in generating structurally correct database queries. However, the slight discrepancies, such as Gemini 1.5-pro coming in first on llm-as-judge preference rankings, but fifth on AST, point to a key difference between technical correctness (how accurately the query matches a reference structure) and user preference (readability, clarity, or "perceived helpfulness").

This preference ranking mechanism suggests an opportunity for query validation and refinement before execution. Recent works, such as Reflexion prompting \cite{reflexion}, DSPy Assertions \cite{DSPyAssertions}, and SPADE \cite{SPADE}, demonstrate how computational constraints can guide LLMs to automatically refine their outputs through self-correction. A similar approach could be applied to database querying, where low preference scores trigger a retry mechanism with specific feedback about potential issues like schema violations, operator misuse, or unclear query intent. This could help prevent problematic queries from reaching the database while providing targeted improvements for subsequent attempts. This further enables future work on preference optimization \cite{DPO, APO} for Function Calling. We present a sample of queries, ranking, and ranking rationales in Table 9.

\begin{table}[h]
\centering
\begin{tabular}{lcc}
\hline
\textbf{Model} & \textbf{Weighted Rank Score} & \textbf{Ranked 1st (\%)} \\
\hline
Gemini 1.5 Pro & 17170 & 20.6\% \\
GPT-4o mini & 16545 & 29.0\% \\
GPT-4o & 15535 & 19.7\% \\
Command R+ & 11845 & 8.7\% \\
Claude 3.5 Sonnet & 11605 & 4.5\% \\
Gemini 2.0 Flash (exp) & 11490 & 10.0\% \\
Llama 3.1 8B Instruct Turbo & 8305 & 4.2\% \\
Command R7B & 8255 & 3.2\% \\
\hline
\end{tabular}
\vspace{1em}
\caption{Preference Ranking results of 8 LLM generated database queries on the DBGorilla benchmark.}
\end{table}

\section{Ablation Studies}

Our main experiments highlight the performance of different LLMs at querying databases with Function Calling using our proposed tool definition and query operators. We additionally present a series of ablation studies to assess how various experimental factors and emerging schools of thought on Compound AI System design influence performance. We explore the impact of requiring a rationale for each tool call, enabling parallel tool calls, using structured output formats rather than Function Calling, and distributing queries across multiple per-collection tools instead of a single unified tool. Shown in Table 6, we find minimal performance variance across these ablations.

\begin{table}[h]
\centering
\setlength{\tabcolsep}{8pt}
\renewcommand{\arraystretch}{1.2}
\resizebox{0.7\textwidth}{!}{
\begin{tabular}{lrr}
\toprule
\textbf{Experiment Type} & \textbf{Exact Match Score} & \textbf{Collection Routing Accuracy} \\
\midrule
Original & 71.8\% & 96.5\%  \\
Original + Tool Rationale & 73.2\% & 96.8\% \\
Original + Parallel Tool Calls Enabled & 71.2\% & 95.9\% \\
One Tool per Collection & 72.3\% & 96.8\% \\
Structured Generation & 72.8\% & 97.1\% \\
\bottomrule
\end{tabular}
}
\vspace{1em}
\caption{Performance comparison across different ablation experiments with GPT-4o.}
\label{tab:ablation_results}
\end{table}

\subsection{Tool Rationale}

We begin by introducing a required \textbf{rationale} argument to our database querying tool. This achieves an Exact Match score of 73.2\% with 96.8\% collection routing accuracy. While the addition of rationales enhances human interpretability for system debugging, its potential benefit for LLM response parsing remains untested. An illustrative example demonstrates how rationales can reveal model misconceptions. When processing the query "How many different types of exhibit highlights are featured in each museum, grouped by museum name?", we observed:

{\textbf{Ground truth:}} Museums, TextAggregation(exhibitHighlights:COUNT), GroupBy(museumName)

{\textbf{Predicted:}} Museums, TextAggregation(exhibitHighlights:TYPE), GroupBy(museumName)

The model's rationale for this query was: \textbf{To determine the variety of exhibit highlights featured in each museum, I will query the 'Museums' collection and perform a frequency analysis on the 'exhibitHighlights' property}. This reveals an interesting misconception where it equates the \textbf{TYPE} aggregation as a frequency analysis operator. This insight suggests opportunities for improving the model's understanding of aggregation and other query operators.

\subsection{Parallel Function Calls}
Another key aspect of Function Calling, illustrated in Figure 4, is the use of parallel tool calls. At each step, the LLM can be allowed to make simultaneous tool calls. In this ablation, we set \textbf{parallel\_tool\_calls} to be true and score each query based on the highest scoring tool called. This achieves a slightly lower Exact Match score of 71.2\% with 95.9\% collection routing accuracy. With parallel tool calls enabled, GPT-4o averages 1.21 calls per query. Upon inspecting the parallel tool calls, we find that this typically results in calls to complementary collections such as a query for the \textbf{Restaurants} collection and \textbf{Reservations} collection in the use case shown in Table 1. In our discussion section, we present a further analysis of how parallel function calls may impact Compound AI System design.

\subsection{One Tool per Collection}
We begin by decomposing our tool definition into a tool per collection. This has important implications for scaling given the token limit for tool descriptions imposed by the LLM providers tested in this study. Splitting collection across multiple tools provides a practical solution for systems with a large number of collections. In our experiments, the one tool per collection approach achieved an Exact Match score of 72.3\% with 96.8\% collection routing accuracy, demonstrating performance comparable to other implementations. We hypothesize that future Compound AI Systems may use the natural language command as input to only create tools for potentially useful collections, or transform multiple collections into a materialized view for Function Calling.

\subsection{Structured Generation}
Our final ablation study challenges the potential bias in LLMs towards their specific Function Calling SDK. We replace Function Calling with Structured Generation using the \textbf{ResponseOrFunctionCall} model shown in Appendix A. Structured generation of function calls achieved a similar Exact Match score of 72.8\% with a collection routing accuracy of 97.1\%. The result suggest that models can effectively work with alternative interfaces for calling external functions. However, this experiment does not eliminate the potential bias towards the Function Calling SDK when processing longer sequences of function calls and their responses. 

\section{Discussion and Future Work}

\subsection{Database Gyms}
A key advantage of synthetic database environments, or database gyms \cite{dbgyms},  is the control offered over schema complexity, such as the number of collections and their property type distributions, as well as query patterns. Real-world databases often contain numerous collections with complex relationships, but publicly available datasets are limited or subject to confidentiality concerns. Synthetic data generation allows us to vary schema sizes, property types, naming conventions, and data relationships. This approach also supports benchmarking edge cases in data management that are challenging to obtain from real data, such as inconsistent naming schemes, partial null fields, or schema evolutions.

Our current setup uses three collections per use case and four properties per collection. Future versions of DBGorilla can introduce more collections, variance in property distribution per collection, and explicit relationships between collections, such as foreign keys. This would enable more sophisticated queries and enable testing for deeper reasoning about interrelated data. Furthermore, generating multiple commands per query-operator combination and introducing more abstract or multi-hop queries would better mimic real-world information needs. This expansion could also include iterative querying scenarios where the result of one query informs subsequent ones. Another opportunity to make the evaluation more robust is to generate multiple queries for each combination of query components, rather than a single query per combination. For instance, if a query involves a text filter and a boolean aggregation, we might produce multiple variations of the natural language command or scenario.

Another avenue for improving the DBGorilla dataset is broadening the function set itself \cite{goex}, extending beyond database querying to incorporate tools such as web search, data visualization, and external analytics platforms. In such a setting, the LLM would need to select among multiple specialized tools based on user intent, possibly orchestrating dependencies between tools (e.g., retrieving data from a database and then creating a chart). This raises fresh design and optimization questions, including how to reliably route requests, how to handle partial results or errors, and how best to refine queries.

\subsection{Querying Databases with Compound AI Systems}
Recent efforts such as Reflexion prompting \cite{reflexion}, DSPy Assertions \cite{DSPyAssertions}, SPADE \cite{SPADE}, and Network of Networks \cite{NoNs} show how LLMs can automatically refine problematic outputs through self-correction steps. Applying similar ideas to database querying could enable an iterative process wherein queries are revised based on validation or user feedback. Recent work has explored more systematic approaches to developing and optimizing LLM pipelines \cite{specifications}, such as DSPy which introduces a programming model that abstracts LLM pipelines as text transformation graphs with declarative modules that can be automatically optimized \cite{dsp, dspy}. Approaches such as MIPRO \cite{MIPRO} or AvaTaR \cite{avatar} could further optimize prompt design and function definitions by contrasting successful and unsuccessful samples, ensuring models learn to manage tokens effectively for large-scale schemas.

\section{Conclusion}
This work demonstrates that Function Calling provides an effective and generalizable interface for enabling natural language database access. Through comprehensive evaluation of 8 LLMs across 5 model families, we show that leading models can achieve high accuracy in translating natural language to structured database operations, with Claude 3.5 Sonnet reaching an Exact Match score of 74.3\% and GPT-4o achieving 71.8\%. Our analysis reveals particular strengths in boolean operations across all models, suggesting a promising direction for optimizing database schemas around boolean properties. The DBGorilla benchmark, with its synthetic schema generation and comprehensive query evaluation framework, provides a foundation for future research in this area. As database systems continue to evolve toward natural language interfaces, Function Calling provides a promising foundation for bridging the gap between human intent and database operations.

\section{Acknowledgements}

We thank Matei Zaharia, Jared Quincy Davis, and the organizers of the Compound AI Systems workshop. We additionally thank Shishir G. Patil, Joseph E. Gonzalez, and the organizers of Sky Camp. For helpful conversations, we thank Liana Patel, Omar Khattab, Krista Opsahl-Ong, Arnav Singhvi, Isaac Miller, Thomas Ahle, Herumb Shandilya, Charlie Cheng-Jie Ji, Sarah Wooders, Charles Packer, Shirley Wu, Devin Petersohn, Augustas Skaburskas, Sebastian Neira Farriol, John Trengrove, Sebastian Witalec, JP Hwang, and Jonathan Tuite.

\bibliographystyle{unsrt}  
\bibliography{references}  

\clearpage
\appendix
\section{Primary Tool Schema Tested}
The OpenAI interface for defining tools to be used with Function Calling contains a string-valued \textbf{type} of the tool and a JSON-valued \textbf[function]. The nested \textbf{function} has a string-valued \textbf{name} of the function, another string-valued \textbf{description} of what the tool does, a list of strings-valued \textbf{required} arguments signaling which of the \textbf{parameters} are required for the tool call and finally, another nested JSON-valued \textbf{parameters} for controlling the tool. The nested \textbf{parameters} further have a string-valued \textbf{type} of the argument such as string, integer, or boolean, followed by a JSON-valued \textbf{properties}.

\begin{lstlisting}[
    language=Python,
    basicstyle=\small,
    numbers=left,
    breaklines=true,
    frame=single
]
query_database_tool = {
  "type": "function",
  "function": {
    "name": "query_database",
    "description": f"Query a database with an optional search query or optional filters or aggregations on the results.\n\nIMPORTANT! Please be mindful of the available query APIs you can use such as search queries, filters, aggregations, and groupby!\n\nAvailable collections in this database:\n{collections_description}",
    "parameters": {
      "type": "object",
      "properties": {
        "collection_name": {
          "type": "string",
          "description": "The collection to query.",
          "enum": collections_list
        },
        "search_query": {
          "type": "string",
          "description": "A search query to return objects from a search index."
        },
        "integer_property_filter": {
          "type": "object",
          "description": "Filter numeric properties using comparison operators.",
          "properties": {
            "property_name": { "type": "string" },
            "operator": { "type": "string", "enum": ["=", "<", ">", "<=", ">="] },
            "value": { "type": "number" }
          }
        },
        "text_property_filter": {
          "type": "object",
          "description": "Filter text properties using equality or LIKE operators",
          "properties": {
            "property_name": { "type": "string" },
            "operator": { "type": "string", "enum": ["=", "LIKE"] },
            "value": { "type": "string" }
          }
        },
        "boolean_property_filter": {
          "type": "object",
          "description": "Filter boolean properties using equality operators",
          "properties": {
            "property_name": { "type": "string" },
            "operator": { "type": "string", "enum": ["=", "!="] },
            "value": { "type": "boolean" }
          }
        },
        "integer_property_aggregation": {
          "type": "object",
          "description": "Aggregate numeric properties using statistical functions",
          "properties": {
            "property_name": { "type": "string" },
            "metrics": {
              "type": "string",
              "enum": ["COUNT", "TYPE", "MIN", "MAX", "MEAN", "MEDIAN", "MODE", "SUM"]
            }
          }
        },
        "text_property_aggregation": {
          "type": "object",
          "description": "Aggregate text properties using frequency analysis",
          "properties": {
            "property_name": { "type": "string" },
            "metrics": {
              "type": "string",
              "enum": ["COUNT", "TYPE", "TOP_OCCURRENCES"]
            },
            "top_occurrences_limit": { "type": "integer" }
          }
        },
        "boolean_property_aggregation": {
          "type": "object",
          "description": "Aggregate boolean properties using statistical functions",
          "properties": {
            "property_name": { "type": "string" },
            "metrics": {
              "type": "string",
              "enum": [
                "COUNT",
                "TYPE",
                "TOTAL_TRUE",
                "TOTAL_FALSE",
                "PERCENTAGE_TRUE",
                "PERCENTAGE_FALSE"
              ]
            }
          }
        },
        "groupby_property": {
          "type": "string",
          "description": "Group the results by a property."
        }
      },
      "required": ["collection_name"]
    }
  }
}
\end{lstlisting}

\clearpage

\begin{lstlisting}[
    language=Python,
    basicstyle=\small,
    numbers=left,
    breaklines=true,
    frame=single
]
class ToolArguments(BaseModel):
    collection_name: str
    search_query: Optional[str] = None
    integer_property_filter: Optional[IntPropertyFilter] = None
    text_property_filter: Optional[TextPropertyFilter] = None
    boolean_property_filter: Optional[BooleanPropertyFilter] = None
    integer_property_aggregation: Optional[IntAggregation] = None
    text_property_aggregation: Optional[TextAggregation] = None
    boolean_property_aggregation: Optional[BooleanAggregation] = None
    groupby_property: Optional[str] = None

class ToolCall(BaseModel):
    function_name: str
    arguments: ToolArguments

class ResponseOrToolCall(BaseModel):
    tool_rationale: Optional[str] = Field(
        default=None,
        description="A rationale regarding whether tool calls are needed."
    )
    use_tools: bool
    response: Optional[str] = None
    tool_calls: Optional[List[ToolCall]] = None
\end{lstlisting}

\section{Additional Query Visualization}
We present a visualization to helper readers further understand the synthetic database schemas and use cases. Table 1 illustrates the 3 collections created in the synthetic Restaurant use case. Each collection contains 4 properties, 2 text, 1 numeric, and 1 boolean. Table 5 further visualizes examples of queries grouped by whether they use text, integer, boolean, or no aggregations. This visualization tool helps overcome the challenge of manually inspecting schemas, a current challenge for inspecting these datasets with CSVs.
\clearpage

\begin{figure*}
    \centering
    \includegraphics[width=1.0\linewidth]{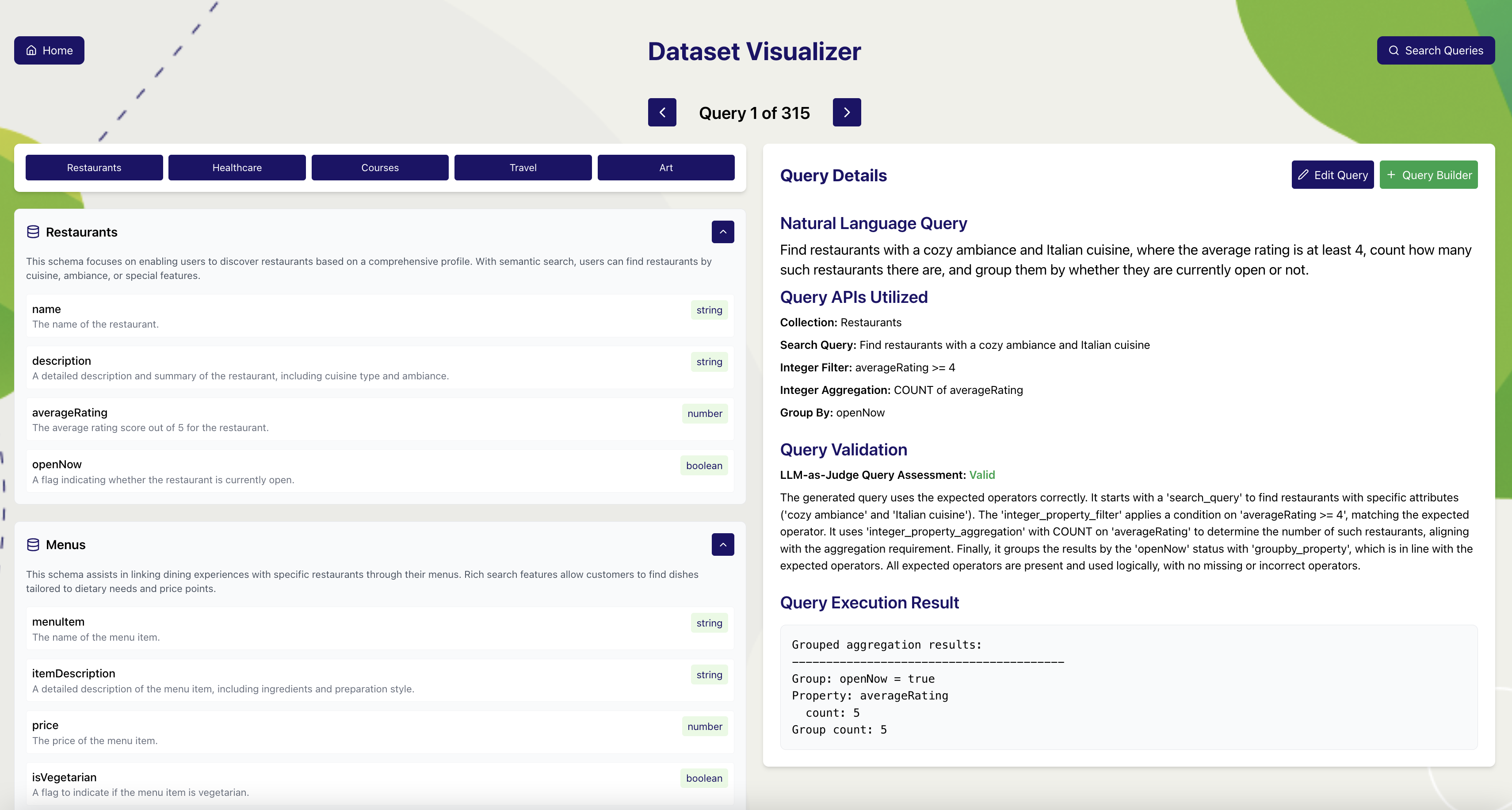}
    \caption{Visualization tool for manually inspecting synthetic query quality and results.}
    \label{fig:enter-label}
\end{figure*}

\begin{table}[htbp]
\centering
\setlength{\tabcolsep}{6pt}
\renewcommand{\arraystretch}{1.2}
\resizebox{\textwidth}{!}{%
\begin{tabular}{p{5cm}p{5cm}p{5cm}p{5cm}}
\toprule
\textbf{Text Property Aggregations} & \textbf{Integer Property Aggregations} & \textbf{Boolean Property Aggregations} & \textbf{No Aggregations} \\
\midrule
Find all Italian restaurants with a cozy ambiance and an average rating of 3.5 or below. Group them by whether they are open, and aggregate the most common words in their descriptions. & What is the average price of seasonal specialty menu items under \$20, grouped by whether they are vegetarian or not? & What are the most highly-rated vegan-friendly brunch spots that are currently open, and can you provide a breakdown of these spots by cuisine type? & What romantic dining locations have an average rating greater than 4.5, and can you group them by whether they are currently open? \\
Find restaurants with a romantic dinner setting and outdoor seating that have an average rating greater than 4. Aggregate the top 5 most mentioned cuisines. & What is the average price of vegetarian healthy salads offered by different restaurants? & How many romantic restaurants with a relaxing atmosphere are currently open and have an average rating of at least 4? & What are some affordable vegetarian dishes that cost less than \$15? \\
Find live jazz music restaurants that are currently open, suitable for a romantic dinner. Group by cuisine style. & What is the average rating of open restaurants with a cozy ambiance, categorized by cuisine type? & Find romantic Italian restaurants that offer organic options and group them by average rating. Show how many are currently open. & Find cozy Italian restaurants that are currently open and group the results by their average rating. \\
How many romantic Italian restaurants with vegan options and a rating above 4.5 are there? Show examples of their descriptions. & What is the average party size for reservations with more than 5 people, grouped by whether the reservation is confirmed? & What percentage of restaurants known for romantic dining settings are currently open, and how are they grouped by average ratings? & Show me open restaurants with a romantic ambiance and group the results by their average rating. \\
How many restaurants are currently open and known for a cozy atmosphere, categorized by cuisine? & What is the average price of affordable vegetarian meals with healthy ingredients, grouped by restaurant? & How many Italian restaurants are currently open? & Find trendy restaurants with a cozy atmosphere and group them by whether they are currently open or not. \\
What are some cozy restaurants that are currently open? Summarize the most common types of cuisine. & What is the highest average rating among currently open restaurants with excellent ambiance and food quality, whose names start with 'La'? & Find Asian restaurants with a cozy ambiance. Determine what percentage are open, and group open restaurants by average rating. & Find restaurants characterized by a cozy ambiance suitable for an intimate dinner, that are currently open and have an average rating of at least 4 stars. \\
Can you find cozy Italian restaurants with a romantic ambiance and group them by their average rating? Provide a summary of common features for open restaurants. & What is the average price of all the menu items available across the various restaurants in the system? & How many reservations are there with a party size of 5 or more? Count how many are confirmed and group by party size. & Find all restaurants that have the word 'Cafe' in their name. \\
Which restaurants have a cozy atmosphere and a romantic ambiance? Aggregate the top 5 cuisines overall. & Find clinics that provide orthopedic care and are rated above 4.0 in satisfaction. Group results by whether they are accepting new patients. & For each name under which reservations are made, what percentage are confirmed? & Show me all the vegetarian items on the menu and group them by their name. \\
How many unique menu items are there in the restaurant menus priced under \$20? & & & Which vegetarian menu items are available, and can you group them by their price? \\
\bottomrule
\end{tabular}
}
\vspace{1em}
\caption{A visualization of query samples from the Restaurant synthetic use case categorized by aggregation type.}
\label{tab:query-categorization}
\end{table}

\begin{figure}
    \centering
    \includegraphics[width=1\linewidth]{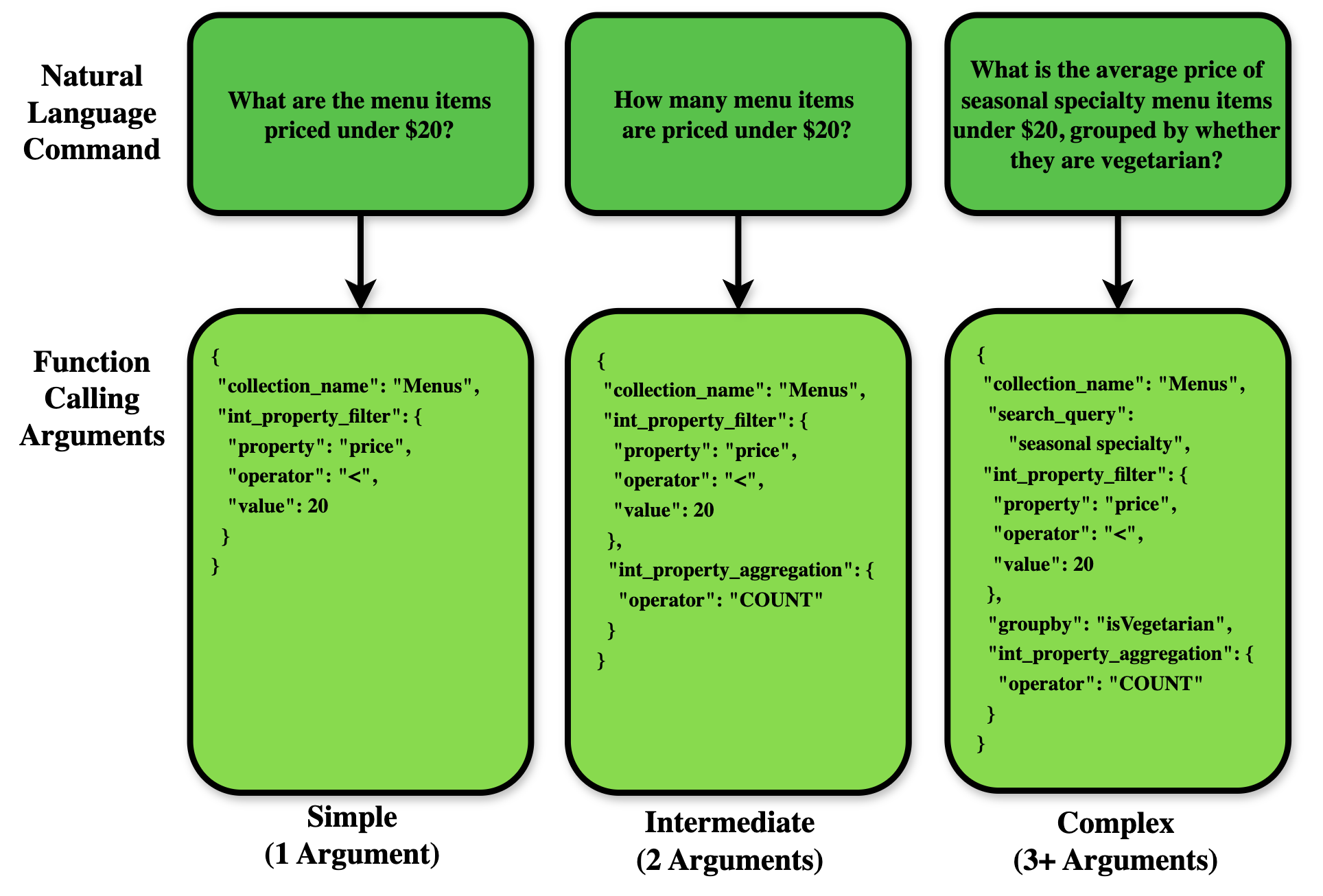}
    \caption{An illustration of natural language commands translated to Function Calling arguments for our proposed tool definition. Natural language command to Function Calling examples are further separated by simple, requiring a single argument, intermediate, requiring two arguments, and complex, requiring three or more arguments.}
    \label{fig:enter-label}
\end{figure}

\clearpage
\section{Performance Analysis by API Component}

\begin{table}[h]
\centering
\setlength{\tabcolsep}{7pt}
\renewcommand{\arraystretch}{1.1}
\resizebox{1.0\textwidth}{!}{%
\begin{tabular}{lrrrrr}
\toprule
\textbf{Component Type} & \textbf{GPT-4o} & \textbf{GPT-4o-mini} & \textbf{Claude 3.5 Sonnet} & \textbf{Command R+} & \textbf{Command R7B} \\
\midrule
Search Queries        & 78.75\% & 79.38\% & 83.75\% & 50.00\% & 38.75\% \\
Integer Filters       & 71.25\% & 86.25\% & 73.75\% & 68.75\% & 35.00\% \\
Text Filters          & 37.50\% & 42.50\% & 46.25\% & 38.75\% & 31.25\% \\
Boolean Filters       & 87.50\% & 86.25\% & 87.50\% & 65.00\% & 42.50\% \\
Integer Aggregations  & 73.75\% & 72.50\% & 73.75\% & 60.00\% & 45.00\% \\
Text Aggregations     & 70.00\% & 66.25\% & 73.75\% & 52.50\% & 35.00\% \\
Boolean Aggregations  & 62.50\% & 72.50\% & 66.25\% & 63.75\% & 31.25\% \\
GroupBy Operations    & 71.70\% & 75.47\% & 72.96\% & 53.46\% & 31.45\% \\
\bottomrule
\end{tabular}
}
\vspace{1em}
\caption{Performance Analysis by API Component (GPT-4o, GPT-4o-mini, Claude 3.5 Sonnet, Command R+, Command R7B)}
\label{tab:component-analysis-1}
\end{table}

\begin{table}[h]
\centering
\setlength{\tabcolsep}{7pt}
\renewcommand{\arraystretch}{1.1}
\resizebox{1.0\textwidth}{!}{%
\begin{tabular}{lrrr}
\toprule
\textbf{Component Type} & \textbf{Gemini 1.5 Pro} & \textbf{Gemini 2.0 Flash} & \textbf{Llama 3.1 8B} \\
\midrule
Search Queries        & 81.25\% & 41.88\% & 52.50\% \\
Integer Filters       & 82.50\% & 46.25\% & 26.25\% \\
Text Filters          & 41.25\% & 25.00\% & 27.50\% \\
Boolean Filters       & 86.25\% & 42.50\% & 32.50\% \\
Integer Aggregations  & 77.50\% & 36.25\% & 32.50\% \\
Text Aggregations     & 70.00\% & 37.50\% & 30.00\% \\
Boolean Aggregations  & 52.50\% & 31.25\% & 30.00\% \\
GroupBy Operations    & 72.33\% & 35.85\% & 23.27\% \\
\bottomrule
\end{tabular}
}
\vspace{1em}
\caption{Performance Analysis by API Component (Gemini 1.5 Pro, Gemini 2.0 Flash, Llama 3.1 8B)}
\label{tab:component-analysis-2}
\end{table}

\clearpage

\section{Preference Ranking Explanations}
\renewcommand{\arraystretch}{1.3}
\setlength{\tabcolsep}{1em}
\begin{longtable}{|p{0.25\textwidth}|p{0.40\textwidth}|p{0.25\textwidth}|}
\caption{A detailed view at preference rankings and their explanations produced by the LLM-as-Judge ranker.}\\
\hline
\textbf{Query} & \textbf{Ranking Explanation} & \textbf{Ranking} \\
\hline
\endfirsthead
\hline
\textbf{Query} & \textbf{Ranking Explanation} & \textbf{Ranking} \\
\hline
\endhead
Which museums that are specifically open today can I visit? & All models except Llama-3.1-8B-Instruct-Turbo correctly used boolean filter. Llama model mistakenly used integer filter. & gpt-4o-mini (1), command-r-plus (2), gpt-4o (3), gemini-1.5-pro (4), gemini-2.0-flash-exp (5), command-r7b (6), claude-3-5-sonnet (7), Llama-3.1-8B-Instruct-Turbo (8) \\
\hline
What is the average entry fee for museums grouped by whether they are open today or not? & Most models correctly grouped by openToday property and calculated mean entry fee. Command-r7b introduced unnecessary boolean aggregation. & gpt-4o-mini (1), command-r-plus (1), gpt-4o (1), gemini-1.5-pro (1), gemini-2.0-flash-exp (1), claude-3-5-sonnet (1), Llama-3.1-8B-Instruct-Turbo (2), command-r7b (3) \\
\hline
What is the total market valuation of all art pieces that are currently on display in the museum? & Models needed to filter displayed pieces and sum valuations. Some models failed to perform correct aggregation or used wrong filter type. & gpt-4o-mini (1), gpt-4o (1), gemini-1.5-pro (1), gemini-2.0-flash-exp (1), claude-3-5-sonnet (1), command-r7b (5), command-r-plus (6), Llama-3.1-8B-Instruct-Turbo (7) \\
\hline
How many different types of exhibit highlights are featured in each museum, grouped by museum name? & Required grouping by museum name and counting distinct types. Best models used both groupby\_property and TYPE metric. & gpt-4o (1), gemini-2.0-flash-exp (2), claude-3-5-sonnet (3), command-r-plus (4), gpt-4o-mini (5), command-r7b (6), gemini-1.5-pro (7), Llama-3.1-8B-Instruct-Turbo (7) \\
\hline
What are the top 3 most frequently mentioned exhibits among all museums, and how many museums are open today? & Required both exhibit identification and open museum counting. Some models missed counting open museums. & gemini-2.0-flash-exp (1), gpt-4o-mini (2), command-r-plus (3), gpt-4o (3), gemini-1.5-pro (3), command-r7b (3), claude-3-5-sonnet (3), Llama-3.1-8B-Instruct-Turbo (5) \\
\hline
What is the percentage of exhibitions currently running grouped by each exhibition title? & Required grouping by exhibitionTitle and calculating PERCENTAGE\_TRUE of currentlyRunning. Some models missed groupby\_property. & command-r-plus (1), gpt-4o (2), claude-3-5-sonnet (3), gpt-4o-mini (4), gemini-1.5-pro (5), Llama-3.1-8B-Instruct-Turbo (5), command-r7b (5), gemini-2.0-flash-exp (5) \\
\hline
What percentage of exhibitions are currently open to the public? & Required calculating percentage using boolean property aggregation. Command-r-plus incorrectly used boolean filter instead. & gpt-4o-mini (1), gpt-4o (1), gemini-2.0-flash-exp (1), command-r7b (1), claude-3-5-sonnet (1), command-r-plus (6), gemini-1.5-pro (7), Llama-3.1-8B-Instruct-Turbo (7) \\
\hline
Which museums open today have notable historical exhibits and how are they grouped by their entry fees? & Required filtering open museums with historical exhibits and grouping by entry fees. Some models missed grouping component. & gpt-4o (1), gemini-1.5-pro (2), claude-3-5-sonnet (3), Llama-3.1-8B-Instruct-Turbo (4), command-r-plus (5), command-r7b (6), gpt-4o-mini (7), gemini-2.0-flash-exp (8) \\
\hline
Find all Italian restaurants with a cozy ambiance and an average rating of 3.5 or below... & Required filtering restaurants, grouping by open status, and aggregating common words. Some models missed text aggregation or grouping. & gemini-1.5-pro (1), claude-3-5-sonnet (2), Llama-3.1-8B-Instruct-Turbo (3), command-r7b (4), gpt-4o (5), command-r-plus (6), gemini-2.0-flash-exp (7), gpt-4o-mini (7) \\
\hline
How many doctors have more than 10 years of experience, and are currently practicing, grouped by their expertise? & Required filtering doctors and grouping by expertise with proper COUNT aggregation. Some models missed COUNT configuration. & gpt-4o (1), gemini-1.5-pro (2), gemini-2.0-flash-exp (3), command-r-plus (4), Llama-3.1-8B-Instruct-Turbo (5), claude-3-5-sonnet (6), command-r7b (7), gpt-4o-mini (8) \\
\hline
\end{longtable}

\end{document}